# MiCull2 – simulating mastitis transmission through milking order

Maya Gussmann[1], Carsten Kirkeby[1], Lars Rönnegård[2]

[1] Department of Veterinary and Animal Sciences, Faculty of Health and Medical Sciences, University of Copenhagen, Frederiksberg, Denmark.

[2] Department of Animal Breeding and Genetics, Swedish University of Agricultural Sciences, Uppsala, Sweden; School of Technology and Business Studies, Dalarna University, Falun, Sweden; The Beijer Laboratory for Animal Science, Swedish University of Agricultural Sciences, Uppsala, Sweden.

## Abstract

Contagious mastitis pathogens can be transmitted through milking. However, previously published simulation models, such as MiCull, have not directly taken this into account. We have reimplemented the MiCull model to model transmission of contagious mastitis pathogens through milking in a milking parlor. This additional complexity requires a substantial increase in computations and a need to structure the program code to make it more flexible for future use. The aim of this paper was threefold: First, to implement the new model in a faster programming language; secondly, to describe the new model, in particular transmission of a contagious mastitis pathogen through milking; and thirdly, to compare three different milking order strategies in regards to prevalence and incidence of intramammary infections. For each scenario, 500 herds with 200 cows each were simulated over 10 years. The model was calibrated using available mastitis parameters from the literature. We hypothesized that milking order should have a considerable effect on disease transmission, especially if the infected cows with clinical enter the milking parlor first and thereby have a high risk of infecting the following cows. The milking order scenarios examined were random milking order and milking clinical cases first, or last. Unexpectedly, there were no large differences between these scenarios for reasonably sized infection rates corresponding to a herd with a moderate level of clinical mastitis in the herd. Larger differences are expected to be found in herds with very high infection rates. We have developed a transmission simulation model of mastitis pathogens using a new mode of transmission by milking order. We expect that this new version of MiCull will be useful for both researchers and advisors since it is flexible, can be fitted to various in-herd situations and the computations are fast.

## Introduction

Mastitis continues to be one of the most costly diseases in dairy cattle and impairs animal welfare (Halasa et al., 2007, von Keyserlingk et al., 2009). It is mostly caused by intramammary infections (IMI) with various pathogens (Zadoks et al., 2011). In Denmark, it is also the main indication for antibiotic treatment in adult dairy cattle (DANMAP, 2022, p. 35). To reduce antibiotic consumption, other measures need to be introduced or intensified. Modelling can be used for investigating which prevention or intervention measures are optimal, given a specific situation.

Mastitis pathogens are traditionally classified into contagious or environmental pathogens (Ruegg, 2017), though in more recent years, this clear distinction has been found to also depend on the specific strain (Zadoks et al., 2011). While environmental mastitis is caused by reservoirs in the cow's environment, contagious mastitis is thought to be transmitted through exposure of the teat to bacteria in milk, which can happen during milking (Ruegg, 2017). However, even though contagious mastitis pathogens are thought to be transmitted through milking, previously published models have not directly taken this into account. They usually assume homogeneous mixing with equal probability for all cows (e.g., Halasa et al., 2009; Gussmann et al., 2018). New knowledge is accumulating with use of strain typing of mastitis pathogens (Woudstra, 2023) and we may therefore expect more detailed knowledge of transmission routes inside dairy barns in the future. Furthermore, cows do not enter the milking parlor in random order (Hansson & Woudstra, 2024), and consequently all cows do not have the same infection probability as assumed in homogeneous mixing. Future modelling of mastitis transmission in dairy barns and its effect on long-term production and levels of clinical mastitis in a herd requires more detailed modelling of individual cows rather than treating the herd as a homogeneous group of animals.

On top of that, to reduce antibiotic consumption, other measures need to be investigated. These could be additional tests before a treatment, increased hygiene due to specific measures, or milking cows in a specific order to minimize transmission between cows. However, measures working with milking order, such as for example a specific milking order of cows or additional disinfection between milkings, cannot be implemented in models that do not model this order. Therefore, to be able to simulate measures taken directly in the milking parlor, we have re-implemented our previously described MiCull model (Gussmann et al., 2018) to model transmission of contagious mastitis pathogens through milking. This would also allow us, in the future, to simulate behavioural dynamics, such as cows following each other into the milking parlor. However, this additional complexity in the model also results in an increase in the required computations and a need to structure the program code to make it more flexible for future needs.

The aim of this paper was therefore threefold: First, to implement the new model in a faster programming language; secondly, to describe the new model, in particular transmission of a contagious mastitis pathogen through milking; and thirdly, to compare three different milking order strategies in regards to prevalence and incidence of intramammary infections.

## Materials and Methods

The MiCull2 model is a re-implementation of the MiCull model (Gussmann et al., 2018), using the Julia Programming Language (Bezanson et al., 2017) and the Julia modules CSV, Distributions (Besançon et al., 2021), Random, StatsBase, and Tables. The model code is available upon request. The earlier version of the model, MiCull, was implemented in R (R Core Team, 2024) and took two to three days to run 500 iterations of one model scenario. Julia, on the other hand, was designed to run fast. Furthermore, it allows working with reproducible environments, meaning that the exact same Julia environment can be recreated, even across different machines. The re-implementation of the model in Julia also allowed us to change the model to a modular structure to allow for easier adjustment in the future.

MiCull2 models a standard Danish dairy herd and transmission of different pathogens causing IMI in the herd on a daily basis. The MiCull2 model is built with modularity in mind, to allow for easier future addition or changes to the model. It is divided into submodules, which focus on different aspects of the modelled dairy herd. The submodules themselves are largely independent of each other and they are merged into one model through linking types and functions that collect input from the submodules and puts them together for the user, adjusting output from other submodules where appropriate.

### Base Submodule

In the base submodule, we model the herd population. This module contains information about all cows in the herd, particularly their status or life stage (calf, heifer, inseminated heifer, pregnant heifer, cow, inseminated cow, pregnant cow, dry cow) and parity. The model is flexible to simulate any number of cows.

An animal keeps their status for a number of days (normally distributed with set means and a standard deviation of 2, see Table 1), after which the status changes automatically to the next one (see above), except for not pregnant heifers and cows. For these, we model whether heat is detected or not with a certain probability (Table 1). If heat is detected, they are moved to either inseminated (but not pregnant) or pregnant status (Table 1). Upon calving, if the calf is female, a new animal with the calf status is added to the herd.

The base submodule also contains basic (economic) culling. Once a week, the model checks whether the number of cows (lactating and dry) exceeds the desired number of cows. If there are more cows than the farmer aims to have, a number of cows are culled. This can happen due to not modelled reasons (with a certain probability, see Table 1), or otherwise according to a culling priority that is determined by reproduction issues (higher culling priority with more inseminations in the current lactation, see Table 1) and parity of the cow (Table 1). However, pregnant cows close to dry-off are exempt from any economic culling.

| Parameter | Value | Source |
|---|---|---|
| Adult cows | 200 | - |
| Milking stalls | 12 | - |
| Days in life stage | | Kirkeby et al. (2016) |
|   Calf* | 365 | |
|   Heifer | 110 | |
|   Inseminated heifer | 41 | |
|   Pregnant heifer | 280 | |
|   Cow | 40 | |
|   Inseminated cow | 41 | |
|   Pregnant cow | 224 | |
|   Dry cow | 56 | |
| Heat detection probability | | Kirkeby et al. (2016) |
|   Heifer | 0.6 | |
|   Cow | 0.36 | |
| Probability for pregnancy after insemination | | Kirkeby et al. (2016) |
|   Heifer | 0.55 | |
|   Cow | 0.42 | |
| Probability to be culled for not modelled reasons | 0.5 | Kirkeby et al. (2016) |
| Culling weights for modelled reasons | | Kirkeby et al. (2016) |
|   Parity 1 | 0.9 | |
|   Parity 2 | 0.1 | |
|   Parity 3 | 0.5 | |
|   Parity 4 | 0.9 | |
|   Parity 5 | 2 | |
|   Parity 6+ | 1 | |
|   Reproduction issues | (Inseminations-1)/6 | Maximum number of inseminations taken from Kirkeby et al. (2016) |

Table 1. Default base model parameters. All parameters can be easily changed to custom values. Only parameters marked by * are not adjustable.

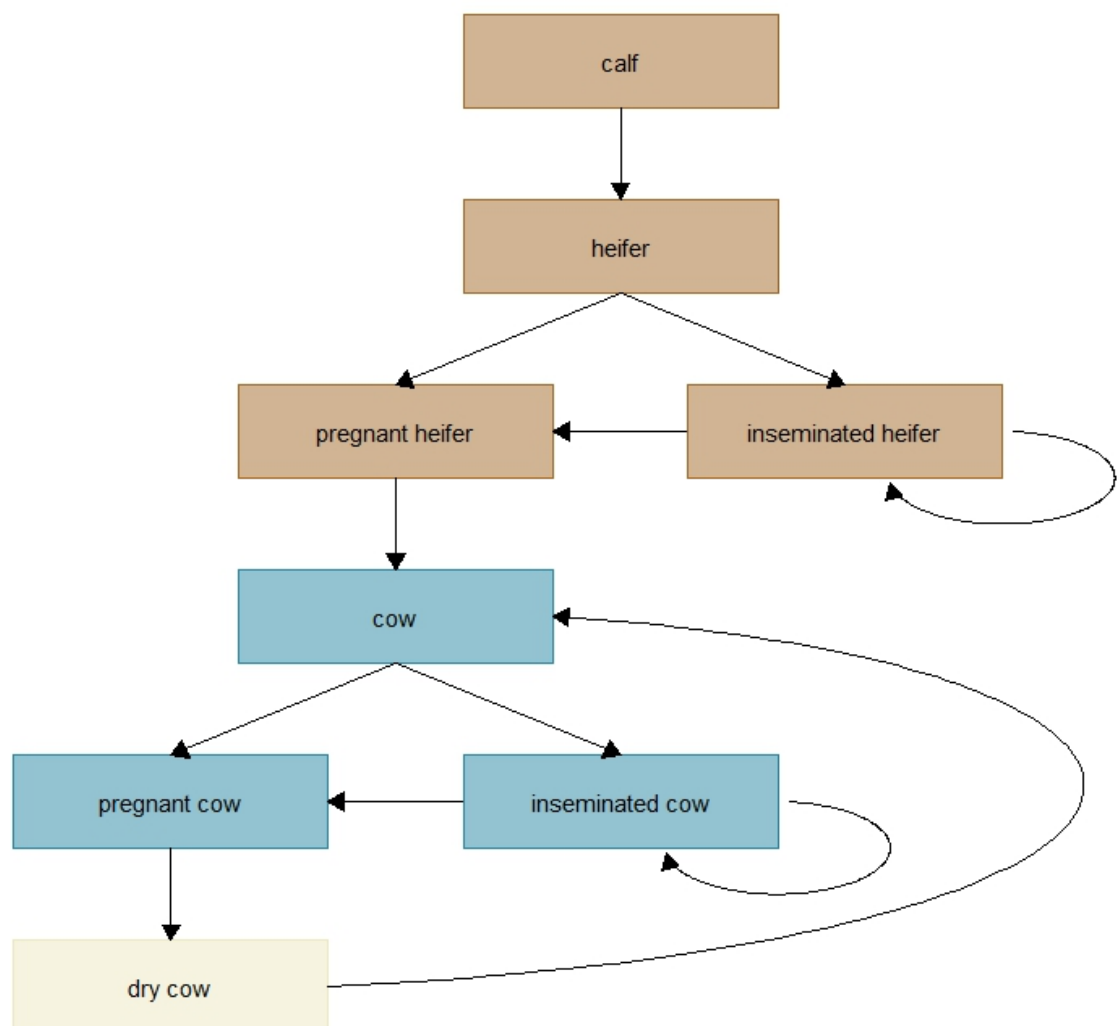

Figure 1. Flow chart depicting how animals can change their lifecycle status. Contagious transmission in the milking parlor involves only lactating cows (blue boxes).

## Mastitis Submodule

The mastitis submodule is a transmission framework for IMI in the herd. Each quarter of a cow is modelled and can become infected independently by one IMI pathogen. As such, each quarter has an IMI status of either susceptible, subclinical, clinical (mild), clinical (moderate), or clinical (severe). At model start, the mastitis submodule is initialized with a 0.25 cow-level prevalence (Table 2). All cows infected on initialization will have one random quarter infected with subclinical mastitis. Each day, there is a probability for subclinical cases to flare up and change to one of the clinical statuses, and a probability for a case to spontaneously recover and become susceptible (Table 2). Clinical cases receive antibiotic treatment for 3 days and return to subclinical or susceptible after treatment ends (Table 2).

Transmission of IMI happens during milking (contagious transmission) and is modelled differently than in the original MiCull model, which assumed homogeneous mixing – as such, infection probability depended on the number of infected quarters for all susceptible quarters (Gussmann et al., 2018). Here, we model the milking order, that is, we model in which specific milking stall and order cows are milked: a teat cup that has been attached to an infected quarter has a probability of transmitting the corresponding pathogen. Up to 5 quarters (i.e., on 5 consecutive cows) that are milked with the same teat cup after an infected quarter have a probability to become infected (subclinical or clinical) with decreasing infection probability ($p_{inf}^n$ where n goes from 1 to 5, see Table 2 for $p_{inf}$). The milking order is random by default, but can be set, as described below. Transmission from one quarter to another of the same cow or from one milking stall to another is currently not modelled. The infection probability given by the milking order is adjusted by a susceptibility factor for every quarter, according to the cow's parity and the quarter's infection history (Table 2).

| Parameter | Value | Source |
| --- | --- | --- |
| Starting prevalence | 0.25 | - |
| Infection probability | 0.01 | Calibration |
| Probability for infection status | | |
|   Subclinical | 0.88 | Halasa et al. (2009), upper limit |
|   Mild clinical | 0 | Not considered in this study. |
|   Moderate clinical | 0.12 | Halasa et al. (2009), upper limit |
|   Severe clinical | 0 | Not considered in this study. |
| Probability for flare-up to clinical status | 0.0011 | Calibration |
| Probability for spontaneous recovery | 0.0064 | Van den Borne et al. (2010) |
| Probability for recovery after treatment | 0.4 | Steeneveld et al. (2011) |
| Infection risk increase after previous infection | 2.4523 | Zadoks et al. (2001) |
| Infection risk increase for parity | | Zadoks et al. (2001) |
|   Parity 1 | 1 | |
|   Parity 2 | 0.8741 | |
|   Parity 3+ | 1.6294 | |
| Disinfection efficacy | 0.5 | Arbitrary estimate |
| Culling weights for modelled reasons | | Bar et al. (2008), Gussmann et al. (2018) |
|   Parity 1, 1 case | 2 | |
|   Parity 1, 2 cases | 4.31 | |
|   Parity 1, 3 cases | 5.37 | |
|   Parity 2+, 1 case | 1 | |
|   Parity 2+, 2 cases | 1.34 | |
|   Parity 2+, 3 cases | 1.7 | |

Table 1. Default model parameters of the mastitis submodule for *Staphylococcus aureus*. All parameters can be easily changed to custom values.

## Submodule Linking

The base submodule is a fully independent submodule; it can run without input from any other submodules. The mastitis submodule uses parts of the base submodule as input, as information about cows (e.g., parity or lactation status) is needed in the transmission framework, as explained above. However, input from the base submodule is not altered within the mastitis submodule.

Output from both submodules is collected and, where necessary, it is combined. In this version of the MiCull2 model, submodule outputs are combined when calculating culling priorities: base culling priorities (base submodule) are altered by output from the mastitis submodule – clinical mastitis cases in the current lactation increase culling priority (Table 2):

w_culling = (w_parity + w_mastitis) * w_inseminations

## Milking Order Scenarios

We tested different milking order scenarios, that is, different strategies a farmer might employ for milking, as described below.

*Random Order.* Cows enter the milking parlor in random order at every milking without any kind of pre-sorting by the farmer. Cow hierarchy is not considered either. During milking, every place in the milking parlor is occupied except, possibly, in the last group of cows.

*Clinical Cases First/Last.* In these scenarios, cows with clinical IMI and early lactation cows (up to 5 days after calving) are separated from the other cows for milking and milked first (or last, depending on the scenario). If the clinical cases are milked first, the milking liners are disinfected after they are milked (see Table 2 for disinfection efficacy). Within the groups, the milking order is random at every milking.

## Model Output

We ran the model for 500 iterations of approximately ten years (3650 days), which included a burn-in period of 1500 days, i.e. model output is considered starting on day 1501 of the simulation until the end. 500 iterations were found to be sufficient for model convergence (data not shown). We collected a number of output parameters, including prevalence (quarter and cow level), number of cases (clinical and subclinical), number of new infections (clinical and subclinical), number of cows in each parity group, and number of model iterations with pathogen extinction.

## Model Calibration

The model was calibrated on the random milking order scenario. We started model calibration at an infection probability of 0.1 and all other transmission parameters as used in the old model (Gussmann et al., 2018). Then, we calibrated infection probability, flare-up probability, and probability to become clinical upon new infection in turns so that the mean modelled cow level prevalence was approximately 15% and the mean monthly clinical incidence was 2 (see Table 2 for calibrated parameters). This calibration was done within the limits of values found in literature. We used visual inspection of the modelled parity distributions to choose the burn-in period so that parity distributions had stabilized (Figure 2).

## Sensitivity Analysis

We ran different sensitivity analyses for all transmission parameters (Table 3) on all three scenarios. During these sensitivity analyses, only single parameters were down- and up-adjusted (Table 3), while all other parameters stayed at their default value.

| Parameter | 40% | 60% | 80% | default | 120% | 140% | 160% |
|---|---|---|---|---|---|---|---|
| Infection probability | 0.004 | 0.006 | 0.008 | 0.01 | 0.012 | 0.014 | 0.016 |
| Probability to become clinical at time of infection | 0.048 | 0.072 | 0.096 | 0.12 | 0.144 | 0.168 | 0.192 |
| Flare-up probability | 0.00044 | 0.00066 | 0.00088 | 0.0011 | 0.00132 | 0.00154 | 0.00176 |
| Probability for spontaneous recovery | 0.002571 | 0.003857 | 0.005143 | 0.006429 | 0.007714 | 0.009 | 0.010286 |
| Probability for cure after treatment | 0.16 | 0.24 | 0.32 | 0.4 | 0.48 | 0.56 | 0.64 |
| Disinfection efficacy | 0.2 | 0.3 | 0.4 | 0.5 | 0.6 | 0.7 | 0.8 |

Table 3. Transmission parameter values for sensitivity analyses. The default values were adjusted to 80%, 60%, and 40%, as well as 120%, 140%, and 160%.

# Results

## Scenario comparisons

The default scenario (Table 3) with random milking order showed a stable parity distribution after 1500 days (Figure 2). The within-herd prevalence was highly variable over the 500 iterations (Figure 3, mean of 24.05% with a standard deviation (sd) of 8.76 after burn-in, 23 pathogen extinctions). The mean monthly clinical incidence was 3.64 cases (Figure 4).

Results were similar for the other two milking order scenarios: when clinical cases were milked first, results showed a mean prevalence of 22.98% (sd 8.76), 37 pathogen extinctions, and a mean monthly clinical incidence of 3.43 cases (see Figure 3 and 4). When clinical cases were milked last, the mean prevalence was 22.27% (sd 9.04), with 41 pathogen extinctions, and a mean monthly clinical incidence of 3.34 cases (see Figure 3 and 4).

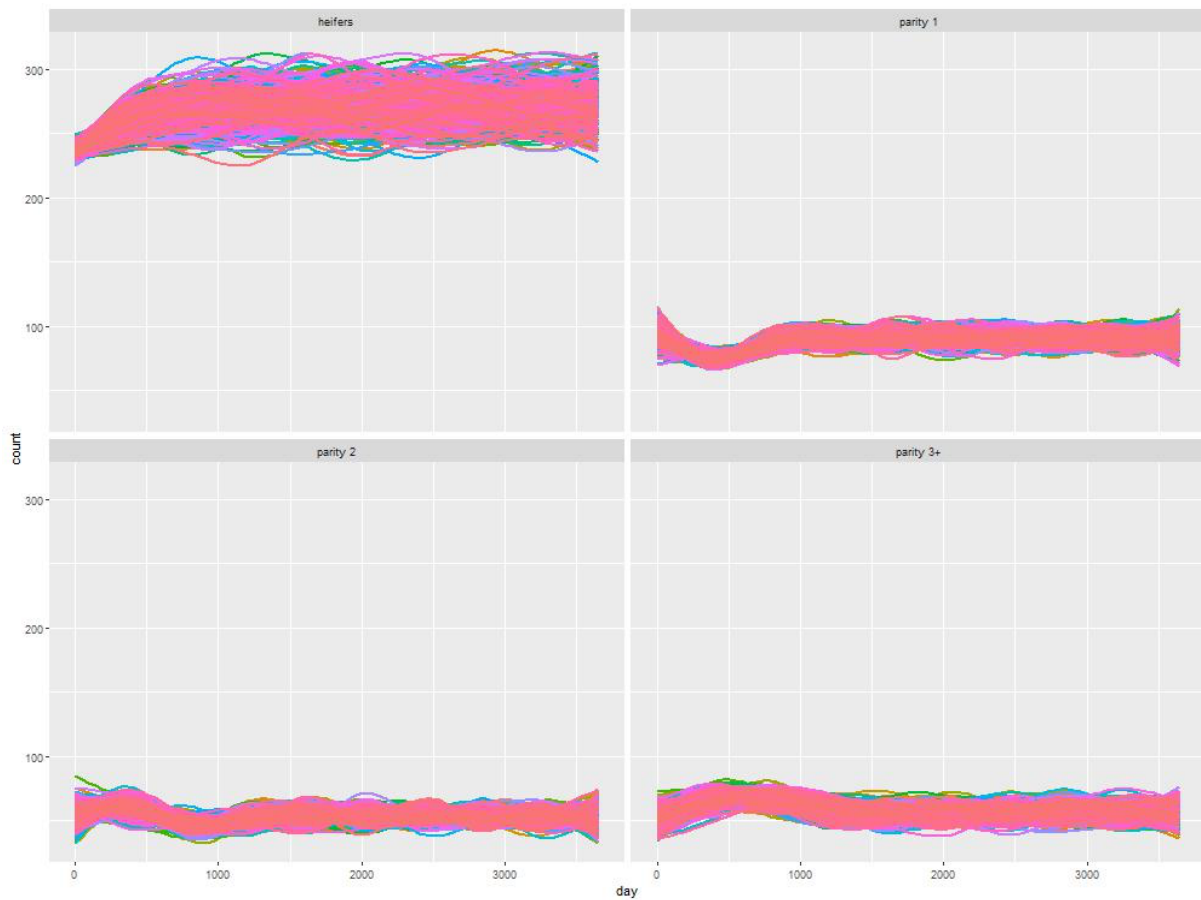

Figure 2. Modelled parity distributions in the default scenario with random milking order with 500 simulation replicates. After approximately 1500 days, the parity distributions seem to be stable.

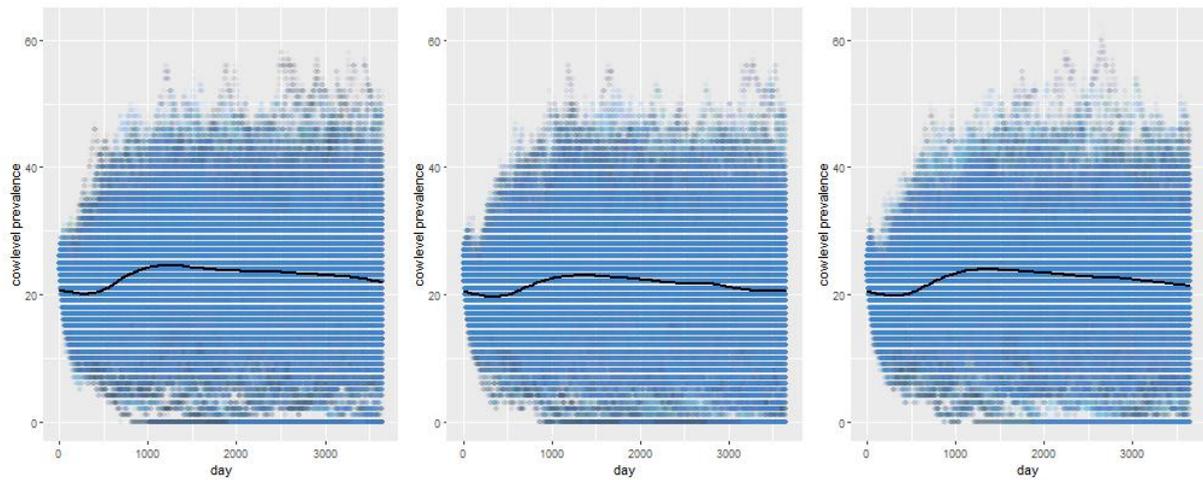

Figure 3. Cow-level prevalence for 500 simulation replicates over 3650 simulated days (blue) with a smoothed mean prevalence (black) for three milking order scenarios (from left to right: random, clinical cases first, clinical cases last).

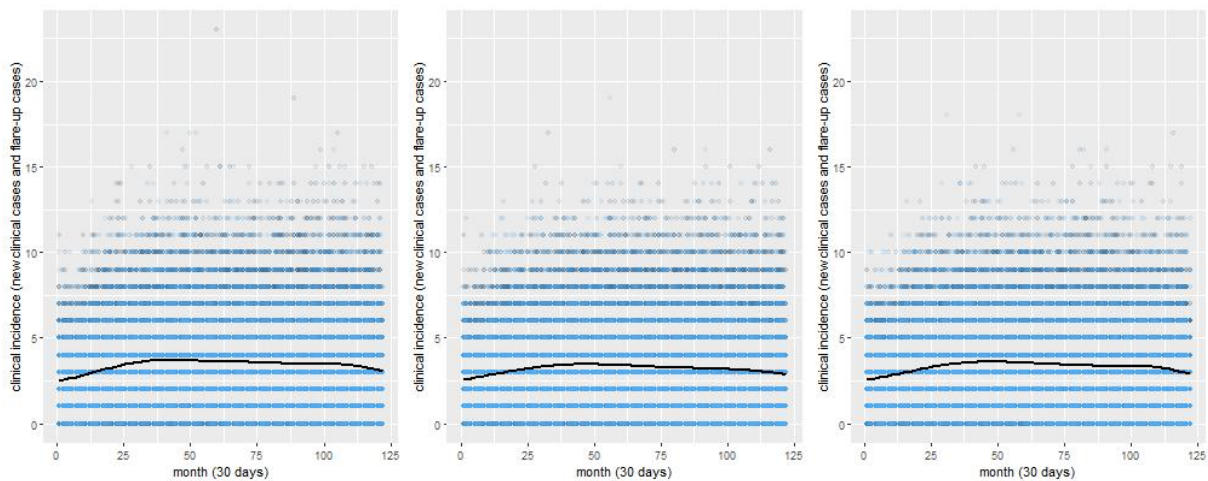

Figure 4. Monthly clinical incidence for 500 model iterations over 3650 simulated days (blue) with a smoothed mean incidence (black) for three milking order scenarios (from left to right: random, clinical cases first, clinical cases last).

## Sensitivity Analysis

Sensitivity analysis using random milking order showed that adjusting the infection probability had a large impact on mean prevalence and the mean number of monthly clinical cases (Table 4), with 493 extinctions at 80% and 500 extinctions at 60% or lower and no extinctions at 120% or higher. The probability for spontaneous recovery had a reversed slightly smaller impact on mean prevalence, with a bit lesser influence on the mean monthly clinical incidence (Table 4). The probability for cure after treatment had a similar influence, but to a lesser effect (Table 4).

Lower and higher probabilities to become clinical (both on infection and by flare-up) led to an increase and decrease in mean prevalence, respectively; with a trend to a lower and higher mean monthly clinical incidence (Table 4).

Increase or decrease of disinfection efficacy had a small influence on respectively decreasing or increasing mean prevalence as well as mean monthly clinical incidence (Table 4).

| Adjusted parameter | 40% | 60% | 80% | default | 120% | 140% | 160% | Output parameter |
| --- | --- | --- | --- | --- | --- | --- | --- | --- |
| Infection probability | 0.00 (0.01) | 0.04 (0.16) | 1.26 (1.9) | 24.05 (8.76) | 54.73 (5.19) | 68.25 (3.08) | 75.69 (2.07) | Prevalence |
| | 0.00 (0.00) | 0.01 (0.02) | 0.17 (0.24) | 3.64 (1.51) | 10.51 (1.56) | 15.02 (1.35) | 18.27 (1.18) | Clinical incidence |
| Probability to become clinical at time of infection | 30.06 (8.65) | 28.15 (8.74) | 26.25 (8.43) | 24.05 (8.76) | 21.82 (8.68) | 19.92 (8.94) | 17.25 (8.23) | Prevalence |
| | 3.28 (1.13) | 3.49 (1.29) | 3.61 (1.34) | 3.64 (1.51) | 3.60 (1.61) | 3.56 (1.76) | 3.31 (1.73) | Clinical incidence |
| Flare-up probability | 29.71 (8.57) | 27.77 (8.69) | 25.79 (8.99) | 24.05 (8.76) | 21.38 (8.51) | 19.24 (8.61) | 17.6 (8.51) | Prevalence |
| | 3.20 (1.09) | 3.42 (1.28) | 3.55 (1.43) | 3.64 (1.51) | 3.51 (1.55) | 3.42 (1.70) | 3.37 (1.78) | Clinical incidence |
| Probability for spontaneous recovery | 71.07 (1.81) | 61.6 (3.14) | 46.38 (6.36) | 24.05 (8.76) | 5.69 (5.39) | 0.85 (1.55) | 0.12 (0.47) | Prevalence |
| | 13.33 (0.68) | 11.04 (0.96) | 7.73 (1.52) | 3.64 (1.51) | 0.83 (0.80) | 0.13 (0.23) | 0.02 (0.07) | Clinical incidence |
| Probability for cure after treatment | 32.28 (8.24) | 29.85 (8.35) | 26.51 (8.68) | 24.05 (8.76) | 19.97 (8.78) | 18.26 (8.33) | 14.91 (8.03) | Prevalence |
| | 5.05 (1.58) | 4.62 (1.55) | 4.04 (1.52) | 3.64 (1.51) | 2.96 (1.44) | 2.70 (1.34) | 2.18 (1.25) | Clinical incidence |
| Disinfection efficacy | 23.18 (8.85) | 23.28 (8.58) | 22.9 (8.64) | 22.51 (8.7) | 22.30 (9.35) | 22.61 (8.87) | 21.97 (8.54) | Prevalence |
| | 3.49 (1.53) | 3.49 (1.49) | 3.42 (1.48) | 3.37 (1.48) | 3.35 (1.58) | 3.39 (1.48) | 3.27 (1.43) | Clinical incidence |

Table 4. Results for sensitivity analyses of the transmission parameters for random milking order. Disinfection efficacy is shown for the milking order where clinical cows were milked first. Shown are the mean cow level prevalences or mean monthly clinical incidences (after burn-in), with standard deviations in parentheses.

# Discussion

This study reinforces that infection probability has the biggest influence on the mastitis situation in a dairy herd. This is not that surprising, given results for infection rates in earlier modelling studies (Gussmann et al., 2018), and the fact that milking hygiene has played a big role in reducing mastitis prevalence since the 1960s (e.g., Neave et al., 1966). Probability for spontaneous recovery was nearly as influential as infection probability, which is easily explained. It removes cases from the herd that, in turn, cannot cause further new cases, thus influencing the infection rate in the herd. Similarly, an increased clinical incidence led to a lower prevalence by potentially removing cases from the herd after a successful treatment. This influence was much lower, as there are a lot less clinical cases than subclinical cases, and cure is also not guaranteed. However, with increased cure probabilities, prevalence was reduced. Surprisingly, the disinfection efficacy had a nearly marginal impact on prevalence. This might be due to only disinfecting after the clinical cows were milked, so transmission through subclinical cases was not impacted at all.

These results are similar to the results we had in our previous model, which assumed homogeneous mixing of cows (Gussmann et al., 2018). The difference in this model is that it is possible to model interventions on the milking and milking order (such as disinfection or milking specific groups together). However, while our model allows for investigation of interventions in the milking parlour, not all type of interventions can be studied at this point, as we do not model horizontal transmission, e.g. through milkers.

The milking order scenarios we examined in this study were random milking order and milking clinical cases first, or last. Before calibration of the model, these scenarios had vastly different results

(results not shown): milking clinical cows separately led to lower prevalences, with the lowest prevalence when they were milked last. However, unexpectedly, after calibration, there does not seem to be a big difference between these scenarios anymore. This is probably due to a really high prevalence with a huge number of clinical cases before calibration. Thus, many cases were milked separately from the "healthy" cows (compared to just a few after calibration), while the ratio of subclinical cases was also lower. In those circumstances, the separation of clinical cases from the rest of the herd seemed to have a large influence on the in-herd prevalence. The interdependence of prevalence and clinical incidence could also be seen during model calibration, when we had to calibrate infection probability and flare-up probability, as well as probability to become clinical upon infection in turns, to model a scenario with a specific prevalence and clinical incidence. These outcomes show that in a herd with a reasonable mastitis prevalence, separating clinical cases from the rest of the herd during milking does not necessarily lead to a better mastitis situation. However, this doesn't mean that farmers should not milk these groups separately anyway, if it makes sense from a management point of view, e.g. clinical cows may already be separated from the rest of the herd anyway. These results just suggest that it doesn't need to be forced, and that there may be other measures the farmer could focus on instead. Furthermore, this study does not allow for any implications for herds with a mastitis problem.

Of course, we have only studied three milking order scenarios, and other scenarios might also be interesting to investigate. Adding to that, our default scenario was a random milking order, while milking order in a real herd is not totally random (e.g., Rathore, 1982). As some of the factors influencing milking order may also be a risk factor for mastitis, it remains unclear if using an "adjusted random milking order" would lead to more visible differences in the outcomes of the three studied scenarios.

In this paper, we decided to not include milk yields, SCC, and economics, as the aim of the study was to describe the new mode of contagious transmission, and to compare the three milking order scenarios in their epidemiological outcomes. Therefore, a module for milk yield was not included at this stage. Furthermore, to keep things simple for the same reason, we did not discriminate clinical mastitis severity in this study, even though it is included in the model. In the future, more targeted interventions like different treatment procedures depending on severity, can also be investigated.

The run time of the model depended on the mastitis situation in the modelled herd, with the low prevalence scenarios running for about 10 minutes, while the high prevalence scenarios in the sensitivity analysis could take up to 2 hours to complete 500 scenario replications. While this means that running various scenarios can still take a bit of time, it is a vast improvement over the previous model, where 500 iterations could take up to 3 days.

Altogether, the MiCull2 model is able to model strain-, cow-, and herd-specific transmission of mastitis pathogens with a daily time step, similar to the original MiCull model. In contrast to the old model or other simulation models of mastitis, the new mode of contagious transmission by milking order allows investigation of interventions in the milking parlor.

# Conclusion

We developed a transmission model of mastitis pathogens using a new mode of transmission by milking order. The model is sensitive to parameter changes in the transmission parameters, particularly infection probability and probability of spontaneous recovery. However, it can be fitted to various in-herd situations. The model does not show a big difference in prevalence, or clinical incidence, between a random milking order scenario, and milking order scenarios where clinical cows are milked first or last.

## Acknowledgements

This project was funded by Formas, a Swedish Research Council for Sustainable Development, Stockholm, Sweden (2019–02276 and 2019–02111).